\documentclass[10pt, conference]{IEEEtran}
\usepackage{cite}
\usepackage{amsmath,amssymb,amsfonts}
\usepackage{algorithmic}
\usepackage{graphicx}
\usepackage{textcomp}
\usepackage{hyperref}
\usepackage{booktabs}  
\usepackage{xcolor}

\def\BibTeX{{\rm B\kern-.05em{\sc i\kern-.025em b}\kern-.08em
    T\kern-.1667em\lower.7ex\hbox{E}\kern-.125emX}}
\begin{document}

\title{CapsuleFS\\
  A Multi-credential DataCapsule Filesystem}

\author{\IEEEauthorblockN{Qingyang Hu}
\IEEEauthorblockA{\textit{Department of EECS} \\
\textit{University of California, Berkeley}\\
Berkeley, CA \\
hqy2000@berkeley.edu}
\and
\IEEEauthorblockN{Yucheng Huang}
\IEEEauthorblockA{\textit{Department of EECS} \\
\textit{University of California, Berkeley}\\
Berkeley, CA \\
hyc@berkeley.edu}
\and
\IEEEauthorblockN{Manshi Yang}
\IEEEauthorblockA{\textit{Department of EECS} \\
\textit{University of California, Berkeley}\\
Berkeley, CA \\
yms@berkeley.edu}
}

\maketitle

\begin{abstract}
CapsuleFS (CFS) \footnote{This project is for both CS 261, Security in Computer Systems and CS 262a, Graduate Computer Systems.} is the first filesystem to integrate multi-credential functionality within a POSIX-compliant framework, utilizing DataCapsule as the storage provider. This innovative system is established based on the Global Data Plane in the area of edge computing. Our comprehensive design and implementation of CFS successfully fulfill the objective of providing a multi-credential Common Access API. The architecture of CFS is methodically segmented into three integral components: Firstly, the DataCapsule server, tasked with the storage, dissemination, and replication of DataCapsules on the edge. Secondly, the middleware, a crucial element running in a Trusted Execution Environment responsible for the enforcement and management of write permissions and requests. Finally, the client component, which manifests as a POSIX-compliant filesystem, is adaptable and operational across many architectures. Experimental evaluations of CFS reveal that, while its read and write performances are comparatively modest, it upholds a high degree of functional correctness. This attribute distinctly positions CFS as a viable candidate for application in real-world software development scenarios. The paper also delineates potential future enhancements, aimed at augmenting the practicality of CFS in the landscape of software development. 
\end{abstract}

\section{Introduction}
Edge computing has seen tremendous development and adoption in recent years. The edge computing paradigm is changing and redefining the boundaries of computing and how and where data is stored and computed. Instead of having a centralized cloud server and delivering data over the Internet to the end user, edge computing brings computing and data storage closer to the user's device, reducing latency and bandwidth, and better preserving the user's privacy.

Global Data Plane (GDP), introduced by Nitesh Mor et al, is an innovative federated storage architecture for edge computing devices \cite{8885071}.  GDP allows users to conveniently access data while it is federatively stored by different storage providers. DataCapsule is the ground truth of GDP, which provides a standardized way to access heterogeneous resources. A DataCapsule is a globally addressable, cohesive encapsulation of data that can live in a widely distributed system. It can provide a unified storage and communication primitive that makes large-scale distributed storage based on DataCapsule possible and feasible. A DataCapsule consists of signed, immutable records that are linked together, and metadata that contains the identity of the DataCapsule and the credentials of its owner.

Due to the DataCapsule paradigm, there are some limitations that need to be addressed to make it more accessible and scalable. First, it would be advantageous to implement a Common Access API (CAPPI) on the client side. This would provide developers with a more user-friendly and concise interface, as the current API necessitates a thorough understanding of DataCapsule. DataCapsule is a blockchain-like data storage unit that requires additional processing before it can be read and written by the client. Second, the single-writer pattern of DataCapsule precludes expanding the current application to a generic file system, due to the possibility of multiple users mounting the same file system and writing to it in a real-world usage scenario. 

In order to facilitate a familiar interface for developers, we utilized Linux's Filesystem in Userspace (FUSE), which is POSIX compliant, as the core of DataCapsule clients. The CFS client operates on the user's device and is responsible for converting the data it receives from the DataCapsule Server into a standard filesystem interface that users can directly use, while utilizing caches and journals to provide excellent performance and crash recovery.

To provide multi-credential support for CFS, we added a middleware layer between the user and the server to manage key distribution, data encryption, signature, and user identity management when it comes to writing data to the file system. The DataCapsule's unique write key is held by the middleware and stored in a Trusted Execution Environment(TEE) for security. The middleware processes the user's write request, encrypts the data with the appropriate keys, and adds a signature to make the change auditable and traceable. In addition, the middleware de-serializes the Access Control List it receives from the DataCapsule server to identify users and protect the DataCapsule server from unauthorized writes.

CFS is mainly developed using Rust and Go. The DataCapsule server and client are implemented in Rust, while the middleware is implemented in Go. Rust offers robust memory safety and thread safety, as well as streamlined integration with other languages while guaranteeing admirable runtime performance. GO's robust and well-documented cryptographic library makes it ideal for middleware implementation, which involves verifying, signing, and forwarding write requests from multiple clients to the server.

Our goals with CFS are twofold:

\begin{itemize}
\item To rigorously adhere to POSIX standards, supporting necessary file and directory operations (creation, deletion, reading, writing, modification) across various operating systems while providing reasonable performance and maintaining consistency in APIs.
\item To incorporate a mechanism for detailed, multi-credential read and write provenance, enabling tracking and auditing of file and directory access and mitigating common threat models.
\end{itemize}


\section{Threat Model}
\label{sec:threat}
To ensure security while providing a multi-credential file system implementation, we considered several threat models, including man-in-the-middle attack, dishonest server, and leaked keys, while designing CFS to address potential vulnerabilities. All these three threat models are possible situations where malicious attackers can gain illegal read access to decrypt data and also write access to put wrong files into the filesystem.

\subsection{Man-in-the-middle attack(MITM)}
\begin{figure}[h]
  \centering
  \includegraphics[width=0.6\columnwidth]{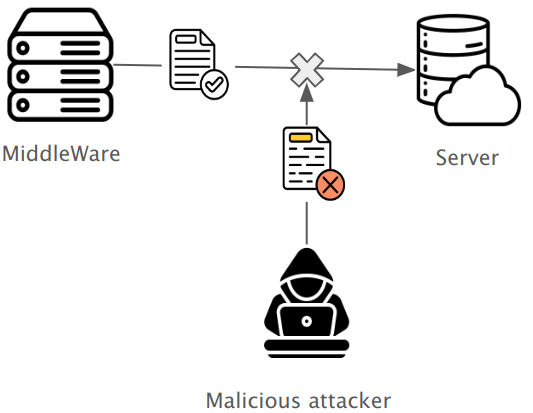}
  \caption{Threat Model: Man-in-the-middle Attack}
  \label{fig:tm-mitm}
\end{figure}

In the threat model for a man-in-the-middle attack, as illustrated in Figure \ref{fig:tm-mitm}, we examine the scenario where middleware requests updated data from the DataCapsule server. During this process, a malicious attacker can impersonate the middleware and substitute the updated file with a counterfeit one, thus illegally writing to the DataCapsule server. Similarly, such attacks could also occur when the client is sending data to the middleware, posing a risk of data interception or alteration.

\subsection{Dishonest server}

\begin{figure}[h]
  \centering
  \includegraphics[width=0.6\columnwidth]{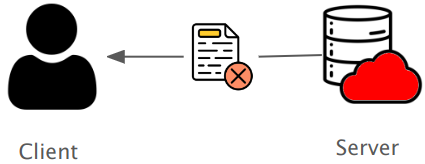}
  \caption{Threat Model: Dishonest Server}
  \label{fig:tm-dh}
\end{figure}

The dishonest server threat model, as shown in Figure \ref{fig:tm-dh}, does not involve malicious attackers. Instead, the service provider is dishonest and wants to serve an illegal file to the client. When the client requests to read from the server, the server may return a false file. This can happen when the service provider is compromised and the storage service is taken over by malicious attackers.

\subsection{Leaked private key}

\begin{figure}[h]
  \centering
  \includegraphics[width=0.6\columnwidth]{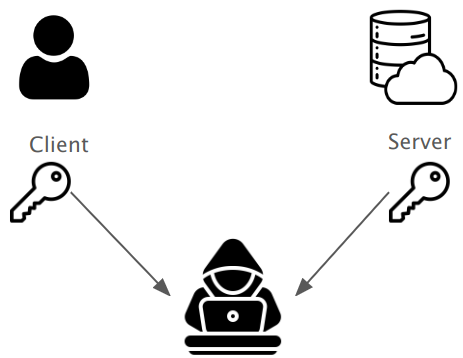}
  \caption{Threat Model: Leaked Key}
  \label{fig:tm-lk}
\end{figure}

In the Leaked Private Key threat model, as depicted in Figure \ref{fig:tm-lk}, both the user's selectively shared read key (or decryption key) and the write key (or encryption key) are susceptible to compromise through attacks. If the user's read key is leaked, a malicious attacker can gain unauthorized access to decrypt data in the file system. Similarly, if the write key is compromised, the server will permit unauthorized writes to the file system, enabling attackers to insert arbitrary data.

\section{Related Work}
\label{sec:related}

\begin{figure}[h]
  \centering
  \includegraphics[width=\columnwidth]{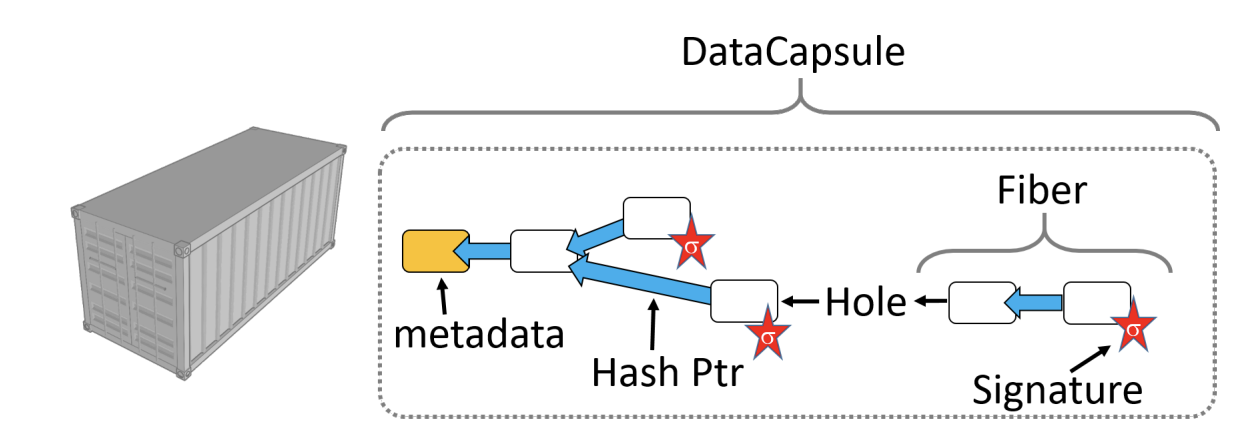}
  \caption{DataCapsule structure\cite{8885071}}
  \label{fig:dc}
\end{figure}
\subsection{Global Data Plane (GDP)} 

As briefly discussed in the introduction, Global Data Plane (GDP), consisting of append-only logs and a routing layer, is designed as a federated edge computing architecture to manage data using DataCapsule as the fundamental collection of data items. \cite{8885071} DataCapsule is represented as a standardized metadata enveloping opaque data transactions as depicted in \autoref{fig:dc}. Each DataCapsule has a unique name derived from hashes over its metadata, enabling it to be globally discoverable. Instead of focusing on infrastructure, GDP offers a ``platform vision'' that allows developers to convey properties, such as performance and security, to the underlying infrastructure to address resource heterogeneity. It provides functionality including a consistent interface, secure storage, secure routing mechanism, administrative boundaries, and locality-awareness etc. It serves as the physical backing for DataCapsule.

\subsection{CapsuleDB} 

To simplify data management on the edge environment using DataCapsule and accelerate data retrieval, a key-value store (KVS), called CapsuleDB, has been proposed. CapsuleDB is the first database and KVS designed for GDP \cite{mullen2022capsuledb}. It provides developers with a straightforward interface that upholds security and can run in a trusted execution environment (TEE), such as Intel SGX, which mitigates attacks from malicious operating systems. To enhance data retrieval speed, an indexing system is utilized to trace active data and leverage the distinct structure of DataCapsules for the natural aging out of older data.

\subsection{IPFS}

The InterPlanetary File System (IPFS) represents a paradigm shift in the domain of distributed storage, characterized by its wholly decentralized architecture \cite{IPFS}. A hallmark of IPFS is its innovative content-addressing scheme, which employs hash-based Content Identifiers (CID). This distinctive approach not only facilitates the decentralization process but also imbues CIDs with inherent self-certification and permanence properties. Notably, IPFS distinguishes itself as the inaugural file system to incorporate a  Merkle Directed Acyclic Graph (MerkleDAG) structure. This structure is instrumental in establishing decentralized trust within its peer-to-peer network, a critical feature for enhancing the integrity and reliability of distributed systems.

After reviewing the current state of the research on DataCapsule, it is apparent that there's a need for a multi-credential filesystem where provenance is provided and data security is guaranteed. Our research aims to fill this gap by introducing CFS as a state-of-the-art filesystem that fits within the GDP framework. This approach not only strengthens data protection but also ensures greater adaptability and scalability to various user roles and data types, making it a versatile solution for modern data management challenges.

\section{Design}
\label{sec:design}
\subsection{Architecture} 
\begin{figure*}[t]
  \centering
  \includegraphics[width=0.8\textwidth]{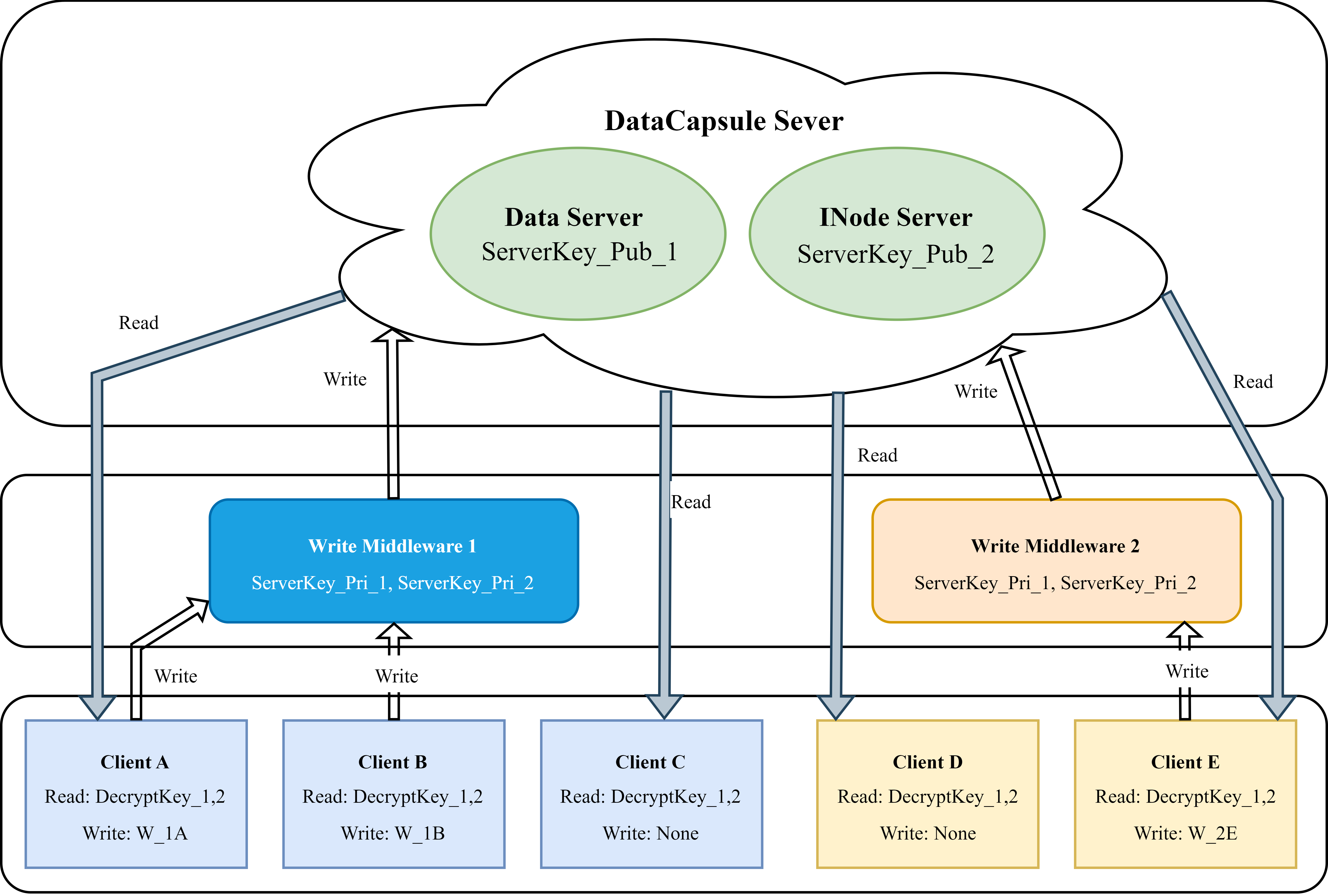}
  \caption{CFS Architecture}
  \label{fig:architecture}
\end{figure*}

The structure of our system, as illustrated in \autoref{fig:architecture}, comprises three distinct components: the DataCapsule Server, the write Middleware, and the CFS client. In this section, we will delve into a detailed discussion of each component, elucidating their individual roles and functionalities within the architecture. This breakdown aims to provide a comprehensive understanding of how each segment contributes to the system's operation and interplays with each other.

\subsubsection{DataCapsule Server} 
The server is primarily in charge of storing DataCapsule blocks and processing read and write requests. For the normal operation of CFS, a minimum of two distinct DataCapsules is essential. The first capsule, noted as the {\it INode capsule}, is dedicated to storing and managing INode information, while the second capsule, noted as the {\it data block capsule}, stores file data separated into multiple blocks. This separation ensures efficient data organization and retrieval and eases our overall system design. On the client side, INode information from the INode capsule is utilized to reconstruct the filesystem structure, and file data is retrieved from the data block capsule.

The diagram in \autoref{fig:architecture} simplifies this structure by illustrating only two distinct servers, storing each of the two capsules mentioned above. However, in practical implementations, additional servers are often deployed to the edge to replicate those capsules. These servers are designed to handle read requests from the clients and write requests originating from the middleware.

\subsubsection{Middleware} 
The middleware occupies a crucial role in our system, mainly facilitating the multi-credential functionality and enforcing write permissions. This component is strategically designed to operate within a Trusted Execution Environment (TEE), for the need to safeguard sensitive operations. Specifically, the middleware is responsible for securely storing the exclusive write keys associated with the DataCapsules and maintaining Access Control List (ACL) information critical for validating write permissions.

In the data flow, each write request is initiated by a client is first routed through the middleware. Here, a verification process takes place. The client's signatures are scrutinized to ensure authenticity and the ACL is consulted to confirm the client's authorization for writing to the designated DataCapsule. Only upon successful completion of these verification steps is the request further processed. The middleware then wraps an additional layer around the request and encrypts the whole block using the DataCapsule's write key before it is forwarded to the server. This layered approach to request handling not only reinforces security but also maintains the integrity and confidentiality of the data flow within the system.

\subsubsection{CFS Client}
On the client side of our architecture, we employ FUSE (Filesystem in Userspace) to create a POSIX-compliant interface, enabling developers to seamlessly mount the CFS onto their personal computers and use it. Upon initialization, the client interacts with two DataCapsules, retrieving data to reconstruct both the filesystem with the local INode information, and files within the filesystem.

Clients are also tasked with managing key cryptographic elements. Each client is uniquely identifiable with its public key, which can be used to validate the origin of the blocks. They must also safeguard their private key, which is used to sign any blocks they would like to submit to the middleware, as well as manage the ACL. Additionally, they need to maintain the read keys of the DataCapsules, which are shared among all clients. User identities within this framework are uniquely defined by a combination of the client’s public key and their local User ID (UID), fostering a secure and distinct identification mechanism.

To enhance performance on the client side, our filesystem features caching and journaling. This cache is designed to optimize read operations by storing frequently accessed data in the memory. Alongside these features, we also included a journaling system to support crash recovery, thereby ensuring data integrity and system resilience in the face of unexpected failures. We have also incorporated the concept of batched requests in our journaling design to improve write efficiency.

\subsection{Block}
\label{subsec:block}
\begin{figure*}[h]
  \centering
  \includegraphics[width=\textwidth]{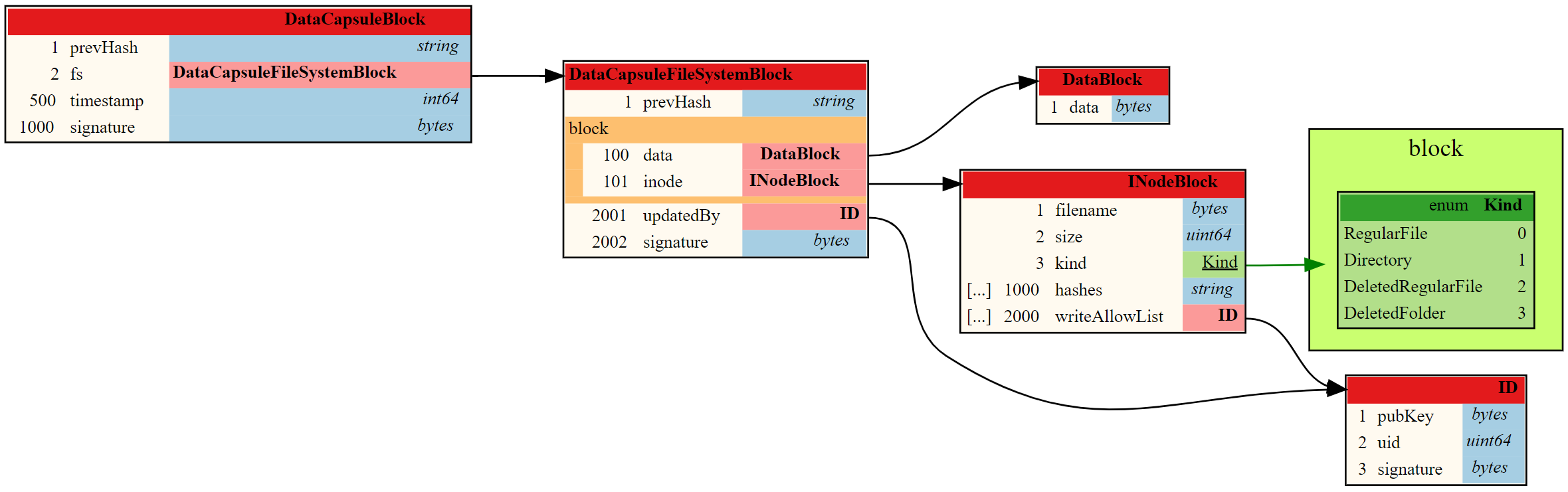}
  \caption{CFS Block Design}
  \label{fig:Middle Block}
\end{figure*}

DataCapsules use blocks for communication and storage. Our design adds additional layers to the DataCapsule blocks, as illustrated in \autoref{fig:Middle Block}. A block can contain either file data or inodes. Within INode blocks, signatures for access control and identity verification are also stored. By design, DataCapsule blocks are linked using hash values, making it easy to verify previous records. During a complete write operation, the client sends the file system block to the middleware along with the client's signature, to prevent any MitM attacks between the client and the middleware (depicted in \autoref{fig:tm-mitm}). The middleware then verifies the block, wraps it around the file system block into a complete DataCapsule block, and sends the encrypted and signed DataCapsule block to the server. This signature is designed to prevent any MitM attacks between the middleware and the server, as shown in \autoref{fig:tm-mitm}.

\subsubsection{Data Capsule Block}

The data capsule block uses the generic structure defined by DataCapsule. It contains a previous block hash to form a chain-like structure, and a file system block as its core.  A timestamp is added by the middleware for conflict resolution. The block is signed and encrypted by the middleware, using the shared write key.

\subsubsection{Capsule File System Block}

Each Capsule File System block contains either an INode block or a data block, determined by whether it is stored in INode capsules or data block capsules. The client who is authoring the block is required to record its ID (public key) and the UID who is initiating the request inside the operating system. The block is signed (but not encrypted) by the client, using the client's signing key.

\subsubsection{INode Block}

The Inode block is used to reconstruct the file system. Each Inode block contains the file name, size, and category. This information is used by the client to convert the structure into the file system. The block also contains ordered links, in the format of hashes, to the data blocks, so that the client could fetch the file data upon request. The ACL list keeps track of all users that are allowed to update the current block, by signing the client's public key and the UID.

\subsubsection{Data Block}

The data block contains the actual file content. It is designed to have a fixed size (for example, 512 bytes), which can be configured during initialization, depending on the workflow.

\subsection{Multi-Credential Management}

In our system, multi-credential management is achieved through a combination of unique identifiers and public key cryptography. Each user's ID is composed of their client's public key and its UID in the operating system, signed by the client's signing key. When a client receives a block, the client first verifies its signature. Then it uses the {\tt updatedBy} field in the INode block to map the UID from the block into the filesystem, but only if the public key matches. If the public key is from a different client, the system maps the UID to {\tt nobody}. This ensures that users are recognized and authenticated based on their unique credentials. Users already included in ACL are permitted to modify that ACL using {\tt chown}, which allows for flexible and secure management of permissions. In the event of a client key leakage, the system can easily revoke access by updating all ACLs associated with that compromised key, ensuring security and mitigating the {\it leaked client keys} threat model shown in \autoref{fig:tm-lk}. Permission enforcement is integrated at the kernel level, with additional checks done at the middleware, restricting write access to only those UIDs listed in the ACL. Furthermore, the ACLs inherit from parent inodes by default, and can be changed independent of others, a practice similar to traditional operating systems. This per-block ACL system enhances security and allows for varied and specific user permissions within the same filesystem, ensuring both flexibility and robustness in access control.

\subsection{Trusted Execution Environment} 
Our middleware is designed to run in a trusted execution enclave, for example, Intel SGX \cite{intelsgx}. Intel SGX is a secure enclave that can provide a protected environment for the code and data \cite{intel_sgx_explained}. 
The reason for using a TEE is for better key protection in our system, to mitigate the {\it leaked server keys} threat model shown in \autoref{fig:tm-lk}. Due to the design and implementation of DataCapsule, the single writer key is the main access control mechanism that prevents the server from illegal writes sent by malicious attackers. In the architecture of CFS and the setting of edge computing, the clients are not secure and trusted. If we distribute the writer key directly to the client, it can be compromised by attackers, and it will take time and resources to distribute new capsules with new keys to the clients. Based on these considerations, we decided to put the writer key in the middleware. 

\subsection{Cache} 

\subsubsection{INode Capsule}
Due to the need for fast access to our filesystem structure, our filesystem features a local, in-memory cache of all the inodes. The local cache maintains 1) a hash-to-inode-number mapping, and 2) an INode to child INode(s) mapping. The cache is rebuilt each time before the filesystem is mounted, by requesting all leaf hashes and fetching their contents. Merkle proofs of such leaves will be validated to ensure the node originated from the authentic DataCapsule, while eliminating the need to recursively fetch all blocks in the path to the root to verify that property, ensuring dishonest servers (discussed in \autoref{fig:tm-dh} will be detected and ignored. Conflicts can be resolved by keeping the latest block written, referencing the \texttt{timestamp} field inside the block. The filesystem will then subscribe to all incoming updates from the INode capsule, updating the local cache using the same conflict resolution strategy.

Our middleware is also aware of the filesystem structure using the same caching strategy, particularly whether an INode has been replaced by a newer INode or not. This is done to prevent leaked keys from being used to submit fraudulent data blocks. 

\subsubsection{Data Block Capsule} 

Our filesystem also caches data blocks fetched from the data block capsule, using the hash as the caching key and employing a hybrid strategy that utilizes both memory and disk. This is done to reduce the write latency for frequently accessed blocks. The cache size can be configured in terms of the number of blocks, with LRU (Least Recently Used) being the default strategy. Subsequent requests for blocks that are cached will be served without querying the server, due to the append-only nature of DataCapsules.

\subsection{Journal} 

Our filesystem utilizes a journal to reduce write latencies. This journal is a queue-like structure, containing blocks to be committed to the middleware. Whenever the filesystem receives write requests, it calculates and manipulates the blocks, then sends them to the journal for commitment. When receiving requests to fetch data blocks, our data cache first checks if the hash is in the journal. If so, data will be returned from the journal without checking the cache or the server. However, as the final hash is only generated by the middleware after it finalizes all signatures, a placeholder hash will be used and replaced once the block is committed. Once all data blocks are committed, the inode block will be updated and sent for commitment. After the inode block is committed, all relevant blocks will be purged from the journal. This significantly reduces write latency as we no longer need to wait for the block to be committed by the server; often, we will need to commit at least two blocks (one data block and one Inode block) for a single write request.

Additionally, the use of a journal provides easy crash recovery. The journal is serialized onto a permanent local storage system, usually hard drives, before the filesystem responds to the write request. In case of failures, the filesystem just needs to read from the journal and resend all blocks to the middleware. Duplication of blocks will not be an issue, as conflicts can be resolved using timestamps. Once all data blocks are committed, we can continue the process mentioned in the previous paragraph to commit the metadata.

To further reduce the number of requests needed for write requests and thus the time needed to fully commit blocks from the journal, as well as to reduce unused blocks on DataCapsules, the journal will batch multiple requests operating on the same block into one. This is done by tracking the INode number and block index associated with each write request. The journal is scanned before sending each block to the server to identify if there are subsequent write requests for the same block. If so, as all subsequent blocks contain the updates from the previous blocks due to the data cache hook, the current block will be dropped, with its hash replaced by the subsequent one.

\subsection{Snapshot}

Our filesystem also allows a user to roll back to a specific timestamp, essentially providing a snapshot / auditing feature. This is achieved by ignoring all blocks written after the specified timestamp while rebuilding the INode cache. This approach offers users easy access to track changes.

\section{Implementation}
\label{sec:impl}

\subsection{Overview} 
We have implemented our file system in Rust, primarily using FUSE, with approximately 550 lines of code. Our filesystem is capable of creating, reading, and writing files and directories. Features like signature verification, block signing, and caching are also implemented. However, due to time constraints, journaling and snapshots were not implemented.

Our middleware is implemented in Go, using approximately 150 lines of code. All features have been implemented, and running inside a Trusted Execution Environment (TEE) will be simulated as we do not have adequate hardware to support this.

Blocks are implemented using Protocol Buffers, and all data exchange is done with gRPC, using HTTP/2 with HTTPS.

We have also created a dummy server, which provides block services to our middleware and filesystem. This is not our focus, so we will not discuss it in our paper.

Our implementation is available on GitHub: \url{https://github.com/hqy2000/cfs}. You may read the project descriptions to learn how to run it in your environment.

\subsubsection{Middleware} 

Due to the major functionality of encryption and communication of the middleware, we decided to implement the middleware using GO. For the encryption part, we used GO's crypto library and used the randomization, encryption, and decryption algorithms including RSA, SHA256, and pkcs1v15. To implement the communication functionality of the middleware, we used GO's gRPC library.

\subsubsection{Client} 

Due to the need for close interaction with the operating system, we have chosen Rust to implement our filesystem. This choice was made because of Rust's superior memory management and debugging experiences when compared to C++. We utilized third-party libraries to aid our implementation; for example, \texttt{tonic} for gRPC and Protocol Buffers, \texttt{rsa} for signature creation and verification, \texttt{fusers} for bridging with the native FUSE interface, etc.

\section{Evaluation} 
\label{sec:eval}
We evaluated CFS's performance in the following aspects:
\begin{itemize}
    \item We measured CFS's read and write performance using a high-resolution timestamp counter of the processor to give an accurate measurement of the system's performance. 
    \item We measured CFS's performance on specific applications to show its potential for software development.
    \item We ran the above two benchmarks on NFS to form a comparison with CFS's performance.
    \item We conducted the simulation of a series of attacks to evaluate if CFS can successfully mitigate the threat models.
\end{itemize}

Our experiments were performed on the above CFS client, middleware, and server implementation. We deployed our CFS implementation, along with NFS, onto a virtual machine with 8 CPUs, 32 GB memory, and 96 GB storage. It is hosted on a bare-metal server with Intel Xeon Platinum 8153 Processor \cite{Intel}, 192 GB DDR4 2133 MHz RAM, and 1.92TB Intel S3610 SSD.

\subsection{Read/Write Performance}
To obtain a comprehensive understanding of the read and write performance of CFS, we measured the latency for various file sizes, recording the average latency for each block. This evaluation included sequential reading of blocks directly from the server, bypassing the client cache, to assess the read performance more accurately. We investigated the influence of cryptographic operations on performance by performing the same experiments with all cryptographic operations turned off. To ensure precision in our measurements, we used the processor's built-in timestamp counter, \texttt{rdtsc}, counting CPU cycles, and utilized inline assembly and system calls for implementation. This approach provided more accurate latency measurements compared to other tools available through high-level libraries. 

\subsubsection{Results}
\begin{figure}[t]
  \includegraphics[width=\columnwidth]{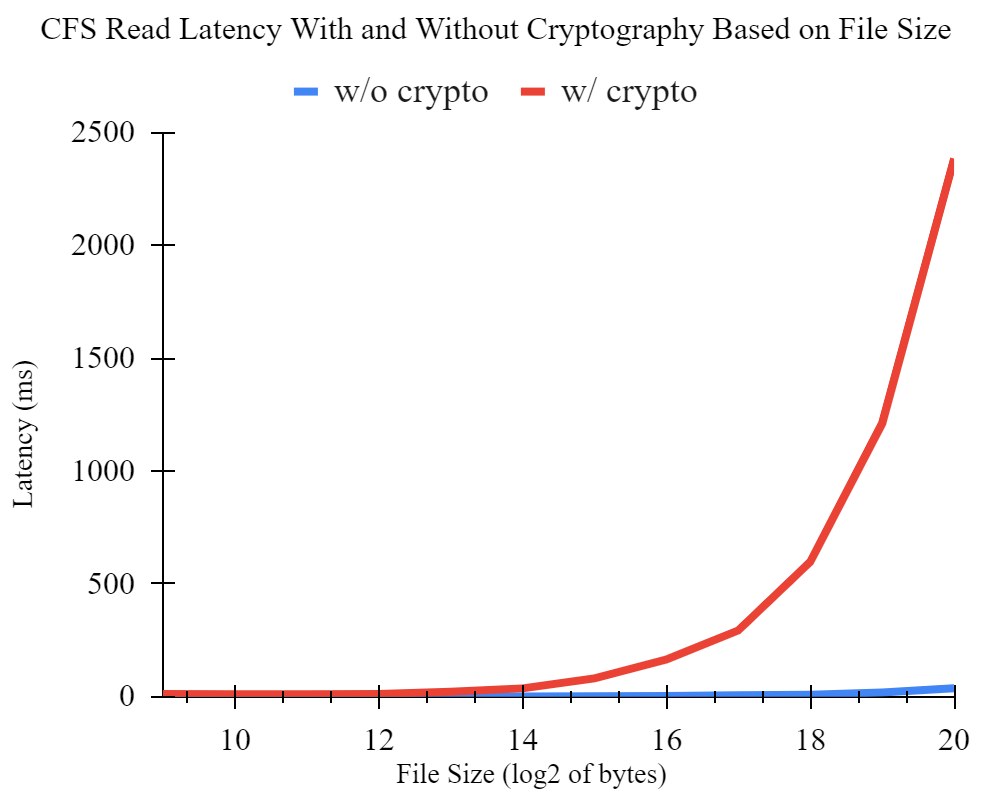}
  \caption{Latencies of read operations based on different file sizes, with and without cryptography}
  \label{fig:CFS read}
\end{figure}

\begin{figure}[t]
  \includegraphics[width=\columnwidth]{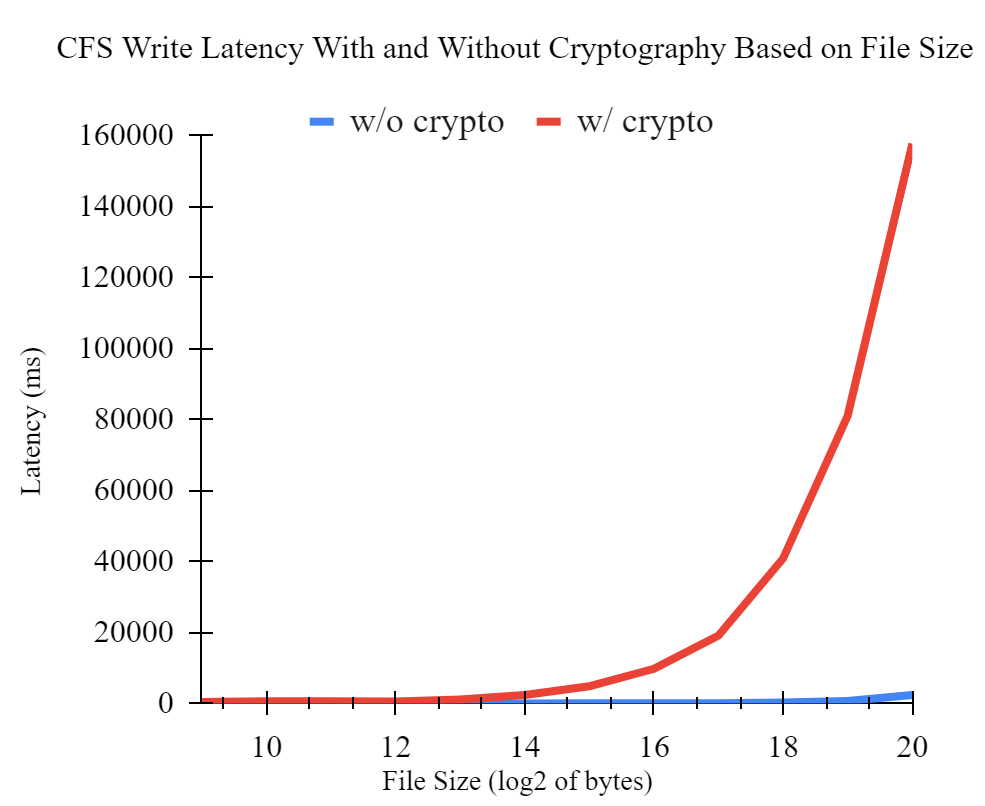}
  \caption{Latencies of write operations based on different file sizes, with and without cryptography}
  \label{fig:CFS write}
\end{figure}

As depicted in \ref{fig:CFS read} and \ref{fig:CFS write}, the read operation outperforms the write operation considerably. This outcome ensues from our utilization of a configurable LRU cache which boosts the read performance by circumventing redundant fetching of recently accessed data from the server. Additionally, because we were unable to implement journaling and batched write requests, the write requests are being processed sequentially by the client and the middleware, resulting in additional time costs. From the graphs, we observe that as the file size doubles, the time required to read and write the file increases exponentially, demonstrating a linear pattern on a logarithmic scale. This indicates that CFS's read and write latency remains uniform for every block.

Another noteworthy aspect is that the cryptographic operations proved to be quite time-consuming. Despite Go's and Rust's well-defined and optimized cryptography libraries, cryptographic tasks still cause significant latency in overall performance. Consequently, write performance is impacted more as the write workflow entails numerous cryptographic operations along the way. Each write request must be signed by the client. Then it passes through the middleware, where the signature and write permission are verified and the request is signed before being sent to the server. In contrast, in the case of a read request, cryptographic operations only occur when the client wishes to decrypt the server's response. 

We conducted measurements of read and write latency for each block without any caches, both with and without cryptographic operations, and the results are presented in \autoref{tab:perblock}. The data demonstrates that enabling cryptography for read operations leads to an increase in latency of approximately 1.2 times compared to when cryptography is disabled. And for write operations, enabling cryptographic operations results in a latency increase of approximately 44 times. Hence an improved efficiency in cryptographic operations will greatly boost CFS's performance.

\begin{center} 
\begin{table}[ht]
    \centering
    \begin{tabular}{c|c|c}
    \toprule
    {\bf Op} & {\bf Latency w/ Crypto}  &  {\bf Latency w/o Crypto} \\
    \midrule
    Read  & 4.13ns & 3.42ns \\
    Write & 411.64ms & 9.09ms  \\
    \bottomrule
    \end{tabular}
    \medskip 
    \caption{Read/write per-block latency (10\% trimmed mean) }
    \label{tab:perblock}
\end{table}
\end{center}

\subsubsection{Comparison with NFS} 

To obtain a comprehensive understanding of CFS's performance in comparison to other distributed file systems, we utilized NFS as a benchmark. Despite NFS's optimization and maturity, we believe that it is a useful indicator in identifying performance bottlenecks and directing our efforts to improve CFS's performance. Here we performed the above experiments in NFS \cite{Ehough}.

Due to the optimized write and read performance in NFS, accurately measuring the sequential read performance presents a challenge. This difficulty arises because it is uncertain whether each read operation is being executed from the disk or the cache, as the prefetch mechanism might influence the process. Regarding the write operations, NFS's server architecture enables concurrent processing of these requests, as opposed to a sequential, one-by-one approach used by CFS. The data in \autoref{tab:nfs_read_write} is derived from the same experiments conducted on a 1MB file. These observations indicate that NFS exhibits significantly smaller latency for both reading and writing operations. Two primary factors contribute to this outcome: firstly, NFS incorporates a prefetch mechanism and supports disk flushing, which means that the experiment does not conclusively determine if the read requests are from the disk or if the write operations are being made to the disk \cite{10.5555/59309.59338}. Another factor influencing CFS's elevated latency is the selection of gRPC for our network stack, which inherently introduces more latency compared to other frameworks like QUIC.

To more precisely compare CFS's read performance with that of NFS, we devised a random read experiment on NFS to mitigate the impacts of prefetching. This experiment entailed performing random reads of 1MB data from various locations within a file so that the entire file could not be prefetched and stored in the cache. Repeated iterations of this experiment yielded an average time of 294.16ms for NFS to randomly read 1MB of data from a 1GB file. Note that the lower read latency in CFS without cryptographic operations is due to the fact that our emulated DataCapsule server stores the data in memory rather than disk for simplicity. Overall, this outcome suggests that prefetching will substantially enhance the sequential read performance of the file system.

\begin{center} 
\begin{table}[h]
    \centering
    \begin{tabular}{c|c|c|c|c}
    \toprule
    {\bf Op} & {\bf CFS w/ crypto} &  {\bf CFS w/o crypto} & {\bf NFS, seq} & {\bf NFS, rand}\\
    \midrule
       Read  & 2385.18ms & 37.88ms & 2.63ms & 294.16ms \\
       Write & 157768.32ms & 2383.16ms & 1.32ms & N/A \\
    \bottomrule
    \end{tabular}
    \medskip 
    \caption{Read/write 1MB file latency}
    \label{tab:nfs_read_write}
\end{table}
\end{center}

\subsection{Application Performance}
Considering that CFS encompasses a comprehensive set of file and directory operations, our objective is to test its viability and efficiency for application in the realm of software development. This evaluation will be conducted through a series of application benchmarks, which are intended to test the capabilities of CFS in a variety of scenarios.

In this study, we selected two distinct tasks frequently employed to evaluate the suitability of a filesystem for software development applications. The initial task involves a compilation process, where the log-structured file system (LFS) benchmark serves as the compilation target \cite{10.1145/146941.146943}.

The second task encompasses the typical operations of compression and decompression tasks which involve intensive read and write and are integral to a multitude of real-world software applications. The chosen subject for this task is the 'strings' package from the Go programming language, which is approximately 188 Kilobytes in size \cite{lang_2023}. This application benchmark aims to provide a comprehensive assessment of the filesystem's performance in scenarios commonly encountered in the software development process. Here we executed the benchmark in both CFS with cryptography turned on and NFS to form a comparison.

\subsubsection{Results} 

The experimental data, as presented in \autoref{tab:application}, provides empirical evidence regarding the operational correctness of CFS. The successful execution of the designated tasks by CFS corroborates its functional integrity. As discussed in the preceding section, it is evident that while CFS exhibits competent performance, there is a notable space for enhancement in its write performance, especially when compared with more established systems such as the NFS. This is particularly observable in write-intensive tasks like decompression, where CFS demonstrates a longer latency. Despite these areas for improvement, the overall performance of CFS in the application benchmark instills confidence regarding its potential applicability in real-world software development scenarios, provided that targeted optimizations are implemented to bolster its efficiency.

\begin{center} 
\begin{table}[ht]
    \centering
    \begin{tabular}{c|c|c}
    \toprule
    {\bf Application} & {\bf CFS w/ crypto} & {\bf NFS} \\
    \midrule
    {\tt make lfs}  & 22.93s & 0.15s \\
    {\tt compress go/strings} & 7.08s & 0.04s \\
    {\tt decompress go/strings} & 39.28s & 0.13s \\
    \bottomrule
    \end{tabular}
    \medskip
    \caption{Application Benchmark Completion Time}
    \label{tab:application}
\end{table}
\end{center}

\subsection{Attacks Simulation}
To verify our design, we conduct a series of simulations of various attacks to evaluate the resilience and effectiveness of CFS in mitigating these threats. This systematic assessment aims to determine the robustness of CFS in the face of diverse and potentially sophisticated attack vectors, primarily the attack scenarios as outlined in the preceding section.

\subsubsection{Man-in-the-middle attack(MITM)}
The resistance of CFS to Man-in-the-middle attacks was evaluated. A malicious attacker's behavior was simulated by replacing the content of a put request. In the first case, only the file data in the put request was replaced. Due to the mismatch of the signature in the request, the put request was identified as invalid and rejected by the middleware. In the second case, we replaced both the file data and the signature. However, since the attacker's ACL key is not in the allowed list, and our client's private key is computationally impossible to guess, the middleware again rejected the request.

\subsubsection{Dishonest server}
We also assessed the CFS client's capability to handle a dishonest server. To simulate a malicious service provider or an honest server that has been compromised by an attacker, we substituted the file data on the server side. When a read request was made, the client verified the merkle proof and the hash value to ensure data consistency. However, due to the server's arbitrary alteration of the data, the client was unable to verify the hash and/or the merkle proof, rendering the attack unsuccessful. 

\subsubsection{Leaked private key}
In this scenario, we revoked a user's decryption key to simulate a situation where the client's sigining key has been compromised. After removing the key from the access control list, the user no longer had write access to the file data associated with the revoked key. All write requests correctly rejected by the middleware, making the attack unsuccessful.

\section{Future Works}
\label{sec:future}

\subsection{Prefetch} 
In earlier discussions regarding the read performance of CFS, we identified the incorporation of a prefetch mechanism as a potential area for enhancement. The fundamental premise of this improvement lies in leveraging the server's capability to handle multiple requests concurrently. By adopting prefetching, CFS can proactively retrieve and store data that is likely to be accessed in the near future, placing it in the cache as a background operation. This proactive data retrieval strategy offers a significant advantage: when users subsequently request this data, it can be delivered directly from the cache \cite{10.5555/59309.59338}. An essential aspect of this mechanism is that prefetching occurs in the background, ensuring that the overall performance of CFS is not adversely affected in scenarios characterized by sparse requests. 

The implementation of prefetching in CFS is expected to markedly enhance read performance, especially in cases involving sequential or pattern-based data access. By utilizing cached prefetched data, this mechanism can significantly expedite data retrieval processes, thereby improving the efficiency and user experience of the CFS.

\subsection{Automatic Cache Policy \& Size} 
In our current design, the cache size for data blocks needs to be manually configured, depending on the workload. This may pose challenges for our end users, requiring them to carefully analyze their applications to optimize performance. For instance, users may need to find a balance between their available memory space and their working set to determine the optimal number of blocks to retain in the cache. In the future, our system should be able to dynamically adjust its cache size based on usage patterns and available memory space.

Additionally, our cache currently utilizes a simple Least Recently Used (LRU) policy. However, for many applications, more complex strategies might be necessary. For example, a Least Frequently Used (LFU) policy may offer better performance in scenarios such as model training, where some datasets are accessed more frequently than others. These datasets may be spread out, rendering the LRU policy less effective in efficiently caching them. In the future, the filesystem should have the capability to dynamically change its cache policies. This enhancement will enable our filesystem to provide a more versatile interface for our end users.

\subsection{Network Stacks} 
Another area for improvement is the choice of network stacks. In our current approach, for simplicity, we use gRPC as a method to exchange block data across our service. gRPC operates over HTTP or HTTPS protocols, which may lead to protocol overhead, such as unnecessary handshakes. In the future, a more refined design focusing on the network protocol may be required. For example, to maintain simplicity, we might consider using the QUIC or HTTP/3 protocol to exchange gRPC requests, thereby reducing handshakes \cite{rfc9000}. Alternatively, we could develop our own protocols based on TCP or UDP to minimize such overhead. Reducing protocol overheads would definitely help to decrease our latencies.

\subsection{DataCapsule Server} 
Currently, we have developed the CFS client, middleware, and server ourselves. Due to time constraints, the functionality and performance of the server part of CFS is not optimized. Samuel Berkun has implemented a data capsule server, dc-prototype \cite{dc-prototype}, which can be used in our file system. Dc-prototype provides a convenient solution and support for routing and connection in the server. In our future work, we will integrate dc-prototype into our server to provide more features and better performance. Moreover, using dc-prototype as a common server can allow other data capsule projects to integrate with CFS.

\subsection{Platform Independent Serialization} 
We utilized gRPC and protobuf for communication between different parts of the system \cite{protobuf}. However, we encountered serialization issues during this process. We discovered that message serialization differs between Go and Rust when using gRPC with protobuf. This resulted in problems with signature verification between the middleware and the server. In the future, a platform-independent serialization method that can be used by both the GO and Rust components of the system will be needed.

\subsection{Sigchain} 
In our design to better deal with key leakage on the user side, we designed a key revoke mechanism that allows a user or the server to invalidate user decryption keys. In the future improvement of this project, Sigchain might be used \cite{blum2022zoom}. Sigchain is the user access control mechanism proposed by Zoom to manage and revoke user-trusted devices. Our project can also use Sigchain to manage and revoke users' read keys, and we expect it to be more secure and efficient than our current implementation, by eliminating the need to track every key in each block.

\section{Conclusion} 
\label{sec:conclusion}
In this paper, we have demonstrated the functionality of CFS as a multi-credential filesystem, conceptualized on the framework of the Global Data Plane (GDP). While our analysis acknowledges certain performance limitations in the current iteration of CFS, we have identified and discussed several strategic areas for potential enhancement, thereby augmenting its viability as a practical application in real-world scenarios. Notwithstanding the prospective improvements aimed at optimizing the performance characteristics of CFS, we anticipate the development and integration of a novel mechanism on the server side to reduce the frequency of cryptographic signing operations. Such a development is expected to facilitate more efficient handling of batched write requests, thereby significantly streamlining the operational efficiency of CFS. This prospective advancement will align CFS more closely with the requirements of real software applications.

\bibliographystyle{IEEEtran}
\bibliography{acmart}

\end{document}